\documentstyle[preprint,eqsecnum,aps]{revtex}
\begin{document}
\def\om{\omega}
\def\P{\Phi}
\def\p{\Phi}
\def\square{\kern1pt\vbox{\hrule height 1.2pt\hbox{\vrule width 1.2pt\hskip 3pt
   \vbox{\vskip 6pt}\hskip 3pt\vrule width 0.6pt}\hrule height 0.6pt}\kern1pt}

\def\lta{\mathrel{\spose{\lower 3pt\hbox{$\mathchar"218$}}
     \raise 2.0pt\hbox{$\mathchar"13C$}}}
\def\gta{\mathrel{\spose{\lower 3pt\hbox{$\mathchar"218$}}
     \raise 2.0pt\hbox{$\mathchar"13E$}}}
\def\spose#1{\hbox to 0pt{#1\hss}}

\draft
\preprint{CITA-94-25}
\title{Kinetic Inflation in Stringy and Other Cosmologies }
\author{Janna Levin }
\address{ Canadian Institute for Theoretical Astrophysics}
\address{Mc Lennan Labs, 60 St. George Street, Toronto, ON}
\maketitle
\begin{abstract}

An inflationary epoch driven by the kinetic
energy density in a dynamical Planck mass is
studied.  In the conformally related Einstein
frame it is easiest to see the demands of
successful inflation cannot be satisfied by kinetic
inflation alone.
Viewed in the original Jordan-Brans-Dicke frame,
the obstacle is manifest as a kind of
graceful exit problem and/or a kind of flatness
problem.  These arguments indicate the
weakness of only the simplest formulation.
{}From them can be gleaned directions
toward successful kinetic inflation.

\vskip10pt

98.80.Hw, 98.80.Cq,04.50.+h

\bigskip

\centerline{submitted June 29, 1994}

\end{abstract}

\narrowtext
\vfill\eject

\section{Prelude}

Recently, there has been interest in a possible
gravity-driven, kinetic inflation.
In the standard inflationary picture \cite{alan}, a potential drives
an era of accelerated expansion.
A remarkable alternative
appears in   any theory
for which the Planck mass is dynamical \cite{un},
such as Jordan-Brans-Dicke theories \cite{jbd} or string
theories.
Due to the direct coupling of the Planck field
to the metric, an acceleration of the
cosmological expansion could result, even in the
absence of a potential.
The cosmic acceleration is driven by the unique
kinetic energy density of the dynamical Planck mass.

In Refs. \cite{un} and \cite{me},
general scalar-tensor
theories of gravity, i.e. general Jordan-Brans-Dicke
(JBD) theories, were
investigated.  It was shown that
the pressure is negative and the
expansion of the universe is accelerated  if
(i) the kinetic coupling parameter evolves with
Planck mass subject to a bound or if (ii)
the kinetic coupling
is a negative constant for a certain branch
of cosmological solutions \cite{me}.
It was shown independently in Ref. \cite{ven}
that
an accelerated cosmic expansion could be
driven by the string dilaton.
The low-energy effective action of string theory is
equivalent to a JBD theory with negative
kinetic coupling parameter. Thus the string dilaton produces
an example of this
general property of JBD theories of gravity.

Since an acceleration of the cosmological expansion
is a fundamental element in inflation, it is natural
to wonder if gravity-driven, kinetic inflation could supplant the
standard potential-driven inflation \cite{un}.
Additional motivation comes from string theories.
Previously, string theories were shown to interrupt
potential dominated inflation \cite{steinhardt}.
The kinetic energy in the dilaton field overwhelmed
the potential energy.  As a result, standard
inflation
could not proceed unhindered.
If the kinetic energy in the Planck field could actually
drive inflation, in lieu of the potential,
string theory would not only be
compatible with inflation but would
actually predict an unusual source of inflation.

However kinetic inflation
stumbled from the outset.
It has been uncovered that inflation could not be exited
properly in string theory
if the lowest order effective
Lagrangian is used \cite{ram}.  Possible remedies to the
graceless exit were suggested and are actively
being pursued.

It might be hoped   that more general JBD theories
would be less problematic.  As argued here, this is not the
case.
A nominal condition for the acceleration to be relevant
for the causal physics of inflation was explored in
Ref. \cite{me}.  In this way, the range of
general scalar-tensor theories was restricted.
Building from these ideas,
the application of the kinetic driven acceleration
to the phenomenology  of inflation is studied more fully
in this paper.
The theory is taken to be of the JBD type with a
general kinetic coupling parameter $\om(m_{pl})$
where $m_{pl}$ is the Planck mass.
The only metric-dilaton coupling is assumed to
be a $\p {\cal R}$ coupling
 where $\p$ is
equal to the Planck mass squared, $\p=m_{pl}^2$, and
${\cal R}$ is the Ricci scalar.
Both the JBD frame  and the conformally
related Einstein frame are investigated.
Obstacles to exiting inflation are encountered
and a brief treatment of flatness is given.

\section{Introduction}

The presence of an acceleration alone does not
ensure the success of an inflationary model.
The universe must inflate enough for a causally
connected region to envelop the extent of our
observable universe.  If this sufficient inflation
condition is satisfied, then the horizon problem
of the standard cosmology is resolved.
The question of sufficient inflation is addressed in the
JBD frame and
in the conformally related Einstein frame in this paper.
It is shown
that successful inflation
requires a positive acceleration of the Einstein frame scale factor
 at some time
prior to today.   Upon conformal
transformation to the Einstein frame, no source for
such an acceleration is apparent.
Unless the acceleration is generated as a result of
some
subtle unforseen effects,
this implies that
a successful execution of the kinetic inflation is impossible
in this simplest $\p{\cal R}$ model.

Though it is simpler to view in the Einstein frame,
any
obstacle to successfully completing inflation
can be seen directly in the JBD frame.
The difficulty
will appear
as a kind of graceful exit problem and/or a kind of flatness
problem.
The cosmological solutions can be broken into two branches.
One branch of solutions, later named D-branch solutions,
will make the universe flatter.
However, it is also shown that
inflation cannot be exited successfully into an expanding phase,
as was already noted for the specific string case in Ref \cite{ram}.
The other branch of solutions, later named X-branch solutions,
may be able to exit into
an expanding phase.  However, this branch encounters a sort
of flatness problem.
Both the graceless exit and the flatness problem can be understood
in terms of the lack of an acceleration in the Einstein frame.

Although the difficulties
discussed  raise concerns,
a kinetic driven inflation is not ruled out.
The obstacle applies to the simplest $\Phi{\cal R}$ model
for which the kinetic driven acceleration can be
transformed away.
Possible ways of
avoiding these obstacles can be suggested.
Effects which may be important include
higher curvature couplings which result, for example,
from the full string action, or massive string modes, or a potential
for the Planck field.
The inclusion of a potential may
return the model to potential dominated inflation or may boost the
kinetic energy so that the kinetic inflation is more effective.
Perhaps, with some persistence the obstacles
may be overcome.

\section{Scalar-Tensor Gravity}
\label{aqua}

As a starting point, the action and resultant
equations of motion for the universe and fields
will be presented in this section.
Some of the results of Ref. \cite{me} are
reproduced in this section for completeness.

The gravitational  action for a general Jordan-Brans-Dicke
theory  is
        \begin{equation}{A[g_{\mu\nu},\Phi]=
        \int d^4 x\sqrt{-g}\left [ {\Phi\over 16 \pi}{\cal R}-{\om(\Phi)\over
          16\pi\Phi}
        g^{\mu \nu}\partial_\mu \Phi
        \partial_{\nu}\Phi
        \right ]\ \ . }
        \label{una}
        \end{equation}
The metric signature $(-,+,+,+)$ was used and  ${\cal R}$ is the scalar
curvature.
Newton's constant $G=\Phi^{-1}$.
The field $\Phi$ is related to the Planck mass through
$\Phi=m_{pl}^2$.
A given theory is specified by choosing the functional form
of $\om(\p)$.
The low energy effective string action has the form
(\ref{una}) with $\om=-1$ and $\p=\exp(-\phi)$ where
$\phi$ is the dilaton:
        \begin{equation}{A_{\rm string}=
        \int d^4 x\sqrt{-g}\ {e^{-\phi}\over 16\pi}\
	\left [ {{\cal R}}
	+
        g^{\mu \nu}\partial_\mu \phi
        \partial_{\nu}\phi
        \right ]\ \ . }
	\label{string}
        \end{equation}
The more general action (\ref{una}) will be used throughout.

It is  assumed that the spatial gradients in $\p$ are negligible
and in general
the universe is homogeneous and isotropic.
The metric is thus
Friedmann-Robertson-Walker (FRW).
The equation of
motion for $\p$ is
         \begin{equation}{
        {\ddot \Phi +3H\dot \Phi=
        -{1 \over (3+2\omega)}
        {d\omega\over d\Phi} \dot \Phi^2 }\ \ ,}
        \label{one}\end{equation}
where $H=\dot a/a$ and $a$ is the scale factor.
For economy of notation
define
        \begin{equation}
  {f(\p)\equiv (1+2\om(\Phi)/3)^{1/2}\ \ .}
    \label{def}
   \end{equation}
The $\p$ equation of motion has as solution
   \begin{equation}
    {\dot \Phi={-C\over a^3 f}\ \ .}
    \label{first}
   \end{equation}
The constant of integration, $C$, can
be positive, negative, or zero.

The equation
of motion for the scale factor $a$, obtained from the
Einstein-like equations, $G_{\mu \nu}=(8\pi/\p)T_{\mu \nu}$, is
        \begin{equation}
        H^2+{\kappa \over a^2}
        =
        -{\dot\Phi\over\Phi} H +{\omega \over 6}\left({\dot \Phi
        \over \Phi}\right)^2   \; \ \ .
        \label{two}
        \end{equation}
Eqn (\ref{two}) can be solved for $H$:
    \begin{equation}
   {H=-{\dot \Phi\over 2\Phi}\left (1\pm f\sqrt{1-Z\kappa}\right) }
        \ \ ,
  \label{usefulone}
   \end{equation}
where $Z$ is the ratio of the curvature term to a kinetic piece,
	\begin{equation}
	Z=
	{{1/a^2}\over {f^2/4}\left ( {\dot \p/ \p}\right)^2}
	=\left (
	a^2\p\over 2 C\right )^2
	\ \ .
	\label{ytwo}
	\end{equation}
When $\kappa$ is taken to be zero, it will be explicitly stated.
Nonzero curvature is considered
in \S \ref{flat}.

If curvature is negligible initially, so that
the metric is taken for illustration to be roughly flat, then
$H$ reduces to
    \begin{equation}
   {H=-{\dot \Phi\over 2\Phi}(1\pm f) }
        \ \ .
  \label{useful}
   \end{equation}
There are two branches  for $H$.
If $f>1$ ($\om >0$), then the Hubble expansion
is positive if the upper sign is chosen for
$\dot \p>0$ and the lower sign is chosen for $\dot \p<0$.
If instead $f<1$ ($\om< 0$), then the universe contracts
when $\dot \p>0$ for
either branch and the universe expands  when $\dot \p<0$
for either branch.

Some useful relations can be obtained for later
reference. First consider the equation of motion
(\ref{two}) rewritten as
        \begin{equation}
        \left (H+{\dot \Phi\over 2\Phi}\right )^2={1\over 4}f^2\left (
        {\dot \Phi\over \Phi}\right )^2 -{\kappa\over a^2}
        .
        \label{alt}
        \end{equation}
Taking the square-root and reexpressing this equation gives
        \begin{equation}
        {d\ln(\Phi a^2)\over dt}=\pm\left (-  f{\dot \Phi\over \Phi}\right )
\sqrt{1-Z\kappa}
        \; .
          \label{pm}
        \end{equation}

The horizon distance can be related to $\Phi$ and $a$ with the
aid of  eqn (\ref{pm}).
Consider the three values of $\kappa$ separately.
First, take $\kappa=0$.
Using (\ref{first}) on the left hand side
of (\ref{pm}) and  integrating over $dt$ gives
        \begin{equation}
        \Phi a^2(1-\delta)=\pm C\int{dt'\over a'}
        \label{named}
                \end{equation}
the constant of integration  is  included in
	\begin{equation}
	\delta\equiv {a_i^2\p_i\over  a^2\p}
	\ \ .
	\label{dell}
	\end{equation}
The subscript $i$ denotes initial values
and $\delta$ is always positive.
{}From (\ref{named}), the particle horizon distance can be deduced.
The distance a photon has travelled since the beginning
of time is defined by the integral $d_{\gamma}=\int dt'/a'$.
Thus from (\ref{named}), it follows that
        \begin{equation}
        d_\gamma={\pm \Phi a^3(1-\delta)\over C}
        \; ,
	\label{fug}
        \end{equation}
or
	   \begin{equation}
	d_\gamma=\pm \left (-{\p \over f\dot \p}
	\right ) (1-\delta)
	\ \ .
	\label{gam}
	\end{equation}

The same procedure as was followed
for the flat case can be used to find the
horizon size for $\kappa\ne 0$.
For $\kappa=+1$, the horizon size
is \begin{equation}
d_{\gamma}=\pm (2 a )\arcsin{\left [{a^2\p\over 2 C}\right ]}
\end{equation}
while for $\kappa=-1$ the horizon size is
\begin{equation}
d_{\gamma}=\pm (2 a) {\rm arcsinh}{\left [ {a^2\p\over 2 C}\right ]}
\ \ .
\end{equation}
Some of these relations will prove useful in the following sections.

\subsection{Cosmic Acceleration}

The acceleration of the scale factor  in a flat universe
is given by \cite{me}
   \begin{equation}
   {{\ddot a\over a}=-{1\over 2}\left ({\dot \Phi\over \Phi}\right )^2 f
        \left [ f\pm1-{df\over d\ln
   \p }{1\over f^2}\right ]\\ .}
   \label{look}
     \end{equation}
In order  for $\ddot a>0$, the following condition must be
satisfied:
    \begin{equation}
    { f\pm1-{df\over d\ln
    \Phi }{1\over f^2}<0 \ \ .}
          \label{condition}
     \end{equation}
The scale factor of the universe will accelerate if:

\indent{(i)
the functional form of $f=(1+2\om(\Phi)/3)^{1/2}$ changes as the
universe evolves
such that it obeys the bound of (\ref{condition}).
The bound can in principle be satisfied for any value
of $\dot \Phi$ and for both branches.
Notice also that $f$ is {\it not} constrained
to be less than 1.}\par
\noindent Alternatively, $\ddot a>0$ if\par

\indent{(ii) a constant but negative
$\om$ ($f<1$) is combined with the lower sign only.
As mentioned below eqn  (\ref{useful}),
this corresponds to an accelerated
{\it expansion} only if $\dot \p<0$.
The particular combination $\om=-1$,
$\dot \p<0$ with the lower sign, was studied
in Refs. \cite{ven} and \cite{ram} in the context of
string theory.}

\section{Sufficient Inflation}
\label{infla}

An acceleration of the scale factor alone by
no means guarantees a resolution of the initial
condition problems of cosmology.
Consider the horizon problem.
In the standard
cosmology,
our observable horizon contains
many regions which were causally
disconnected at earlier times.
Consequently, the
smoothness of the observed universe
would appear to have no causal explanation.
In
the inflationary scenario,
a dynamical
explanation of the large scale
homogeneity and isotropy  is provided.
In standard inflation
a potential energy density
drives an era of accelerated expansion.
During the rapid expansion,
a causally connected region that was
small at the beginning of inflation grew
large enough to contain our observed universe.
Subsequently, entropy was produced
as the energy in the potential converted
to radiation.

The question is, if the potential driven acceleration
is replaced by a kinetic driven acceleration, can the
horizon problem be resolved?  In other words,
can a kinetic driven acceleration blow up
a region causally connected early in the history
of the universe large enough to
encompass our observable universe today.

This requirement of sufficient inflation
can be stated in equations.
The comoving size of  a causally connected region
at some earlier time $t_*$ is defined
by the comoving distance a photon has
travelled since the beginning of time, $d_{\gamma}(t_*) / a_* $.
Today, the extent of the observable
universe is
$\sim H^{-1}_o$.
A causal explanation for the
smoothness of the universe today follows if
the comoving size of the observable universe today fits inside a
comoving region causally connected at $t_*$,
	\begin{equation}
	{{d_{\gamma *} \over a_*} \gta {1 \over H_oa_o}
	\ . \ }
	\label{full}
	\end{equation}
A detailed study of (\ref{full}) is left to appendix B.
In the next section,
this expression will be considered in
terms of conformally transformed variables.

\section{Einstein Frame}
\label{conform}

The condition of sufficient inflation can be studied under
a conformal transformation to the Einstein frame.
In the Einstein frame, the theory of gravity is the usual
Einstein theory with a fundamental Planck scale $M_o=1.2 \times
10^{19}$ GeV.
In the Einstein picture the
FRW universe is filled with an ordinary, minimally coupled scalar field.
There is no
acceleration of the Einstein frame scale factor.
However, it is argued in this section
that the sufficient inflation condition
requires an acceleration some time prior to today in the
Einstein frame.  This indicates that the kinetic acceleration
felt in the Jordan-Brans-Dicke
(JBD) frame cannot lead to a successful inflation
model unless additional effects are invoked.

The conformal transformation on the metric
        \begin{equation}
        g_{\mu \nu}=\Omega^2 \tilde g_{\mu \nu}
        \ \ ,
        \end{equation}
defined by $\Omega=M_o/\Phi^{1/2}$ takes the JBD action into
an Einstein theory.
Under the conformal transformation the action becomes
(see for instance Ref.  \cite{books}),
        \begin{equation}
        A=\int d^4x\sqrt{-\tilde g}\left [
        {M_o^2\over 16 \pi}\tilde {\cal R} -{1\over 2}
        \tilde g^{\mu \nu}\partial_\mu \Psi\partial_\nu\Psi
        \right ]
        \end{equation}
A redefinition of the fields was also performed:
        \begin{equation}
        \Psi
        \equiv M_o{\sqrt{3\over 16\pi}}\int  {(1+2\om/3)^{1/2}\over
	\Phi} d\Phi
        \ \ .
\label{redef}
        \end{equation}
Notice that $\Psi$ is real and the energy
density in the $\Psi$-field, $\rho_\Psi=\dot \Psi^2/2$ is positive
if $\om\ge -3/2$.
The momentum associated with the field, $p_\Psi=\rho_\Psi$,
is always positive.

In addition to the conformal transformation, perform the coordinate
transformation
        \begin{eqnarray}
        d\tilde t&=&\Omega^{-1}dt \\
        \tilde a&=&\Omega^{-1} a
	\label{coot}
        \end{eqnarray}
so that the spacetime interval can be written in the usual FRW form,
        \begin{eqnarray}
        d\tilde s^2&=&\Omega^{-2}ds^2\\
        &=&-
(\Omega^{-1}dt)^2+(\Omega^{-1}a)^2\left [{dr^2\over 1-\kappa r^2}
	+ r^2(d\theta^2 +\sin^2{\theta}d\phi^2) \right ]\\
        &=&- d\tilde t^2+\tilde a^2\left [ {dr^2\over 1-\kappa r^2}
	+ r^2(d\theta^2 +\sin^2{\theta}d\phi^2)
        \right ]
        \; . \label{transtom}
        \end{eqnarray}
The metric is the usual FRW metric with scale factor
$\tilde a\propto \p^{1/2} a$.
Incidentally, Einstein time
always increases with increasing JBD time.
To see this, notice that
the conformal transformation $\Omega=M_o/\p^{1/2}$ is always
positive.  If $dt>0$, so that time ticks forward in the
JBD frame, then $d\tilde t$ is always greater than zero.
Since $d\tilde t/dt=\Omega^{-1}>0$, the slope
of $\tilde t(t)$ is positive.
Therefore Einstein time always ticks forward with JBD
time.

The evolution of the scale factor $\tilde a$ and of $\Psi$ can be found
directly in
the Einstein frame in a flat cosmology.
The Hubble equation is
	  \begin{equation}
	\tilde H^2=\left ({8\pi \over 3 M_o^2}\right )
    \tilde \rho_{\Psi}
			\ \ ,
   \label{htil}
   \end{equation}
where the kinetic energy density in $\Psi$ is
	   \begin{equation}
   \tilde \rho_{\Psi}={1\over 2} \left ({d\Psi\over d\tilde t}\right )^2
   \ \ .
   \end{equation}
As always, an overdot will be used to denote
a derivative with respect to JBD time.  A derivative with
respect to Einstein time will be written out
explicitly.
The $\Psi$ equation of motion is
    \begin{equation}
   {d^2\Psi \over d\tilde t^2}+3\tilde H{d\Psi \over d\tilde t}=0
   \ \ ,
 \end{equation}
which has solution $d\Psi/d\tilde t=-B/\tilde a^3$ where $B$ is
an arbitrary constant of integration.
Using eqn (\ref{redef}), $B$ can be related to the arbitrary constant
$C$ of eqn (\ref{first}),
\begin{equation}
B={C\over M_o}\sqrt{3\over 16\pi}
\ \ .
\label{geecee}
\end{equation}
{}From the arguments of appendix A,
only $\dot \p<0$ is relevant and so only
$d \Psi/d\tilde t<0$ is relevant.
Therefore we shall
hereafter assume $B>0$ ($C>0$).
Using the solution to the $\Psi$ equation of motion in
$\tilde H$ of eqn (\ref{htil}) gives
\begin{equation}
\tilde H=\pm {C\over 2M_o^2}{1\over \tilde a^3}
\ \ .
\label{eek}
\end{equation}
The upper sign corresponds to an eXpansion
and will hereafter be called X-branch solutions.
The lower sign
corresponds to a Decrease in the Einstein frame scale
factor and will hereafter   be called D-branch solutions.

The Einstein frame scale
factor can be found as a function of Einstein time by integrating
(\ref{eek}),
\begin{equation}
\tilde a=\left [\tilde a_i^3\pm {3C\over 2 M_o^2}(\tilde t-\tilde t(t_i))
\right ]^{1/3}
\ \ .
\label{atil}
\end{equation}
For X-branch solutions $\tilde a>\tilde a_i$ and
the scale factor grows.
For the D-branch solutions $\tilde a<\tilde a_i$ and the scale
factor drops.

For completeness,   $\Psi$ can be found as a   function
of Einstein time,
	\begin{equation}
	\Psi=\Psi_i\mp {M_o\over \sqrt{12 \pi}}\ln
	\left[\tilde a_i^3\pm {3C\over 2 M_o^2}(\tilde t-\tilde t(t_i))
	\right ]
	\ \ .
	\end{equation}
Again, the upper sign refers to X-branch
solutions and the lower sign refers to D-branch
solutions.

Clearly, the second time derivative of the scale factor is
always negative,
   \begin{equation}
    {d^2\tilde a \over d\tilde t^2}=-{1\over 2}
	\left ({C\over M_o^2}\right )^2 {1\over \tilde a ^5} \ \ .
	\label{neg}
	\end{equation}
For X-branch solutions the universe expands at a decelerating
rate.  For D-branch solutions,
(\ref{neg}) gets more and more negative.  The universe
contracts at an accelerated rate.

The Einstein frame cosmology seems to know nothing about
$\om(\p)$ and thus does not appear to distinguish
between different scalar-tensor theories.
However, an observer who carries rulers
and clocks must be included in order to
compare events in one frame with events
in a conformally related frame.
Conformally related observers
do agree on the occurence of events though they disagree on the
interpretation of the physics.
Once an observer is included, the rulers
and clocks of that observer can be shown to
scale as functions of the conformal factor.
The different
scalar-tensor theories can then be distinguished
 \cite{glenn}.

Consider a scalar-tensor theory which
leads to an acceleration of the
cosmic expansion in the JBD frame.
The acceleration is conformally transformed away
in the Einstein frame.  Still,
a kinetic driven acceleration
of the scale factor in one frame and a
deceleration of the scale factor in a
conformally related frame can be made consistent.
The acceleration of the JBD scale factor is attributed to
the rate of change of rulers in the Einstein frame
\cite{me}.

While it is true that the effect can be viewed without
contradication
in either the JBD or the Einstein picture,
the sufficient
inflation condition imposes an additional requirement.
The equivalence of the two pictures means that if
eqn (\ref{full}) is
satisfied in one frame, it must be satisfied in all frames.
If, on the other hand, eqn (\ref{full}) cannot be satisfied in one frame,
then it cannot be satisfied in a conformally related frame.
One could try case by case in the JBD frame to hunt
for a solution to the sufficient inflation condition.
However,  by
looking in the Einstein frame we can immediately show
that expression (\ref{full}), transformed accordingly,
cannot be satisfied by a kinetic-driven acceleration alone,
at least not in a $\Phi{\cal R}$ model.
By considering the Einstein picture, it can be predicted
that any attempt to meet condition (\ref{full}) will
fail unless additional effects are conjured up.

Consider the sufficient inflation condition (\ref{full})
	\begin{equation}
	{d_{\gamma *}\over a_*}\gta {1\over H_oa_o}
	\ \ .
	\label{refill}
	\end{equation}
This expression can be rewritten
in terms of
Einstein variables
by conformally transforming both sides.
Consider first the left hand side.
For all times,
the comoving horizon size is the same in both frames
	\begin{equation}
	d_\gamma/a=\int dt/a=\int d\tilde t /\tilde a
	=\tilde d_\gamma/\tilde a
	\ \ .
	\end{equation}
To transform the right hand side of (\ref{refill}) notice that
in general
	\begin{equation}
	\tilde a \tilde H= a (H-\dot \Phi/2\Phi)
	\ \ .
	\label{okay}
	\end{equation}
The right hand side of (\ref{refill}) can be found
by evaluating (\ref{okay}) today,
$\tilde a_o \tilde H_o= a_o (H_o-\dot \Phi_o/2\Phi_o)$.
By today, the Planck mass should be anchored in the
vicinity of $M_o=1.2 \times 10^{19}$ GeV and
the conformal factor should be nearly one.
In other words, by today, $\dot \Phi_o\sim 0$. So, today,
Einstein and JBD quantities are equivalent.
Therefore (\ref{okay}) reduces to
$\tilde H_o\tilde a_o=H_oa_o$.
It follows that in terms of Einstein quantities,
the sufficient inflation condition requires
	\begin{equation}
     	{\tilde d_{\gamma *}\over \tilde a_*}\gta {1\over \tilde
	H_o\tilde a_o}
	\ \ .
	\label{wow}
	\end{equation}
To clarify, eqn (\ref{wow}) is the conformally transformed
version of the sufficient inflation condition imposed in
the JBD frame.  If the Planck mass today was not near
$M_o$, then eqn (\ref{wow}) would not look so similar to the
JBD form.

To evaluate (\ref{wow}) further, one would need to know
the particle horizon distance.
The distance a photon has travelled since $\tilde t(t_i)$ is
	\begin{equation}
	\tilde d_{\gamma}(\tilde t)=\tilde a\int_{\tilde t(t_i)}^{\tilde t(t)}
	{d\tilde t'\over \tilde a(t')}
	= {M_o^2\over C}\tilde a^3\left (| 1-\delta |\right )
	\ \ ,
	\label{dphotil}
	\end{equation}
where, as in eqn (\ref{dell}), $\delta\equiv {\tilde a^2_i / \tilde
a^2}=(a_i^2\p_i)/ (a^2\p)$.
For X-branch solutions $\delta <1$ while for
D-branch solutions $\delta >1$.
Therefore, in terms of $\tilde H$,
	\begin{equation}
	\tilde d_{\gamma}(\tilde t)={|(1-\delta )\tilde H^{-1}(\tilde
	t) |\over 2}\ \ .
	\label{mess}
	\end{equation}

Consider for now solutions for which the
Einstein universe expands, i.e.
X-branch solutions.
With (\ref{mess}) in eqn (\ref{wow}), sufficient inflation demands
	\begin{equation}
	{1\over \tilde H_*\tilde a_*}\gta {1\over \tilde H_o\tilde a_o}
	\ \ ,
	\label{generous}
	\end{equation}
where factors of order 1 were dropped.
Using the definition of $\tilde H$, the above equation
is equivalent to
\begin{equation}
{d \tilde a\over d\tilde t }|_o\gta {d \tilde a\over d\tilde t}|_*
\ \ .
\label{stup}
\end{equation}
Eqn (\ref{stup}) states
 that the expansion rate is greater today than in the past; that is,
the expansion
increases at some time prior to today and thus the Einstein frame scale
factor accelerates at some time prior to today.

For D-branch solutions the universe contracts.
It is even clearer here that an acceleration is
ultimately needed.  Today we live in the Einstein
frame and today the universe expands, $d\tilde a/d\tilde t|_o>0$.
At some time between today and the epoch of unusual behavior, during
which the Einstein frame scale factor contracted, an
acceleration is needed.

If the sufficient inflation condition is met, even the
conformally transformed scale factor must accelerate.
No acceleration of the Einstein scale factor
is manifest in kinetic-driven inflation
(see eqn (\ref{neg})).
If the effect is to be meaningful for inflation,
this contradiction in the    Einstein picture must be
resolved.

Perhaps there is some way around this obstacle.
It may be that
there is some
unforseen subtlety which
rectifies this.  Most likely, additional effects will need
to be posited.  If such additional effects are capable of accelerating
the
Einstein frame scale factor, the kinetic acceleration
may be altogether replaced.
It should be stressed that the above argument is limited
to $\Phi{\cal R}$ theories.  Higher order curvature terms
which cannot be transformed away, for example,
will not be tainted by these results.

The Einstein frame results reveal in one swoop the obstacle to
completing kinetic inflation.
Although this alone is convincing, it is instructive
to consider the problems as they appear in the JBD frame.
For completeness then,
we finish the paper by considering the JBD frame.
The obstacle will appear as
a graceful exit problem
and/or a flatness problem.
We will see that the shortcomings in the JBD frame can
again be attributed to the absence of an acceleration
in the Einstein picture.

\section{Jordan-Brans-Dicke Frame}

There seem to be so many different possible cases to
study.  However, the set of possibilites relevant
for inflation can be paired down considerably.
While the expansion of the universe may be accelerated
by a variety of forms for $\om$ and values of $\dot \Phi$,
only cosmologies for which $\dot \Phi<0$ and $f<1$ ($\om<0$)
can be relevant for inflation.
For a justification of this statement the reader
is refered to appendix A or Ref. \cite{me}
where the arguement first appears.
There are still two branches of cosmological solutions:
X- and D-branches.
That nomenclature is carried over into the JBD frame
discussion.
\footnote{To connect with the terminology of Ref. \cite{ram},
note the X-branch was there called the minus branch and
the D-branch was there called the plus branch.}

\newpage

The content of the two different branches can be summarized
in the following chart:
\vskip 10truept

{\offinterlineskip
\halign{ \vrule # &  \strut\
	\ # \ & \vrule # \cr
\multispan{3}\hrulefill\cr
height2pt & \omit  &   \cr
& $\dot \Phi<0$,    $\om(\Phi)<0$ &\cr
height2pt  & \omit &   \cr
\multispan{3}\hrulefill\cr}}

\vskip 10truept

{\offinterlineskip
\halign{ \vrule # & \strut\ # & \vrule # &
	\ # \ & \vrule # \cr
\multispan{5}\hrulefill\cr
height2pt & \omit & & & \cr
& X-branch (upper sign) & & D-branch (lower sign) &\cr
height2pt & \omit & & &  \cr
\multispan{5}\hrulefill\cr
height2pt & \omit & & & \cr
& $H>0$              & & $H> 0$           & \cr
& $\Phi a^2$ grows   & & $\Phi a^2$ drops & \cr
& $\tilde a$ eXpands & & $\tilde a$ Drops & \cr
height2pt & \omit & & &  \cr
\multispan{5}\hrulefill\cr }}

Only the range of parameters represented in
the above chart are considered.
This should make discussion of  the JBD frame more manageable.

\subsection{Graceful Exit and the D-Branch}
\label{gracie}

Any obstacle to successfully completing
inflation should be encountered directly in the
JBD frame, without reference to the Einstein
frame.  For D-branch solutions the obstacle
takes the form of a graceful exit problem.
The graceless exit is discussed in this section.

It was argued in Ref. \cite{ram}, in the context of
string theory, that the accelerated inflation
could not be exited gracefully.
As it happens, the string model is an example of the more
general acceleration of \S \ref{aqua}.
Specifically, the low energy effective
string action corresponds to
the D-branch solution for $\om=-1$
in the language of this paper.
It is shown here that
all accelerations in a D-branch solution
will suffer from a kind of graceful exit problem,
regardless of the form of $\om(\p)$.
For variable $\om$, the graceful exit problem is
of a new sort.  The accelerated expansion can be
turned off by tuning $\om(\Phi)$.  However, there is another behavior
which cannot be exited.

As can be seen from quick inspection of eqn (\ref{condition}),
the accelerated expansion can easily be exited if
$\om(\Phi)$ is allowed to vary.  However, for the
D-branch, a branch change is needed if the product
$a\Phi ^{1/2}$ is not to drop forever.
Consider the evolution of the product $a\p^{1/2}$  given
by eqn (\ref{pm}).
For $\dot \p<0$ and the D-branch, notice
eqn (\ref{pm}) becomes
	\begin{equation}
	{d\ln(a\p^{1/2})\over dt}=-{f|\dot\p |\over 2\p}=-{C\over 2
	a^3\p}
	\label{rewri}
	\end{equation}
The quantitiy $a\p^{1/2}$ always decreases.
Today by contrast,
no variation in the strength of gravity has been observed.
It follows that
the field $\p$ must be nearly constant today.
Since the universe still expands, the product $a \p^{1/2}$ grows today.
A smooth transition to our universe appears to be prohibited.

To see this another way,
consider eqn (\ref{rewri}) rewritten in terms of $H$,
	\begin{equation}
	H={|\dot \Phi|\over 2 \Phi}(1-f)
	\ \ .
	\end{equation}
If $f$ were to grow in excess of 1, the epoch of accelerated growth
of the scale factor would be exited, as eqn (\ref{condition})
shows.	However, as $f$ exceeds 1, $H$ will pass through
zero and the universe will undergo collapse.\footnote{
When $\om$ is allowed to be negative, the universe
can undergo collapse even for $\kappa=0$.
In a Brans-Dicke universe, for which $\om$ is
a positive constant, this is not possible
\cite{curvey}.  }
Again, the D-branch does not connect smoothly onto our expanding
cosmology.

If other energy densities which do not couple
directly to $\Phi$ are included, then
   \begin{equation}
    {d\ln(a\p^{1/2})\over dt}=- \sqrt{
	 \left ({f\dot\p\over 2\p}\right )^2+{8\pi\rho_{\rm }\over 3\p}}
   	\ \ .
	\label{ouch}
	\end{equation}
 Regardless of the heating mechanism, if
energy is transfered in a simple way from kinetic into
a hot particle bath, the behavior will not turn around.
In order to turn this around, negative quantities must appear
on the
right hand side of (\ref{ouch}).
Attempts were made in \cite{ram} to include negative
potential energy densities which arise naturally in
supersymmetry.
Although the authors found for their case that
a branch change could be induced by a negative potential,
they also found the branch change occured in pairs.  If the
universe began with the D-branch it would insist on
ending up in the D-branch.

In the conformally related Einstein frame,
$\p^{1/2}a\propto \tilde a$
dropping corresponds to  a contracting universe.
The difficulty in effecting the
branch change can be seen in the Einstein frame as the
difficulty in turning the evolution from a
contraction into an expansion.  As argued
in \S \ref{conform}, an acceleration of the Einstein
frame scale factor is needed.

This trap appears at
first glance to affect only D-branch solutions.
There are entire families of
X-branch solutions which do not immediately
run into this obstacle.
For X-branch solutions, $a \Phi^{1/2}$
grows and a smooth connection onto our
universe may be possible.
However, when curvature is included, the X-branch
will tend to turn into a D-branch if $\kappa=+1$.
An attempt to avoid this leads to a kind of flatness
problem.

\subsection{Flatness}
\label{flat}

In the standard cosmology, the
universe
would quickly veer away from a flat appearance,
unless
extraordinary initial
conditions are imposed.
Initially, curvature is unimportant in determining the dynamics
of the scale factor and the universe looks roughly flat.
As the scale factor grows, the curvature term should
quickly come to dominate in the determination
of the standard cosmological evolution since the
curvature term in the
equation of motion scales as $1/a^2$ while the
standard radiation density term scales as $\rho\sim
1/a^4$.  If $\rho$ is to continue to
influence the cosmic dynamics today, then the entropy in radiation
must be enormous.  In terms of the dimensionless entropy,
$S\propto a^3T^3$, the standard cosmology requires $S\gta 10^{90}$.
The need for such a huge value of the
otherwise arbitrary constant entropy is the famed flatness problem.

During any epoch of acceleration, by contrast,
the universe is made flatter.
This can be demonstrated by
considering the equation of motion
  \begin{equation}
	H^2+{\kappa\over a^2}={8\pi\over 3\p}\rho
	\ \ .
	\end{equation}
Define the curvature radius
    \begin{equation}
	\label{curvy}{
      R_{\rm curv}={R(t)\over |\kappa|^{1/2}\  \  }   }
	\ \ .
	\end{equation}
Comparing the scales
$R_{\rm curv}^{-1}$  and
$H$ shows
  \begin{equation}\label{bas}
	{{ R_{\rm curv}^{-1}\over H}={|\kappa|^{1/2}\over \dot a }
\ \ .}
\label{rat}
	\end{equation}
As the universe accelerates,
$\dot a$  grows.
The importance
of curvature  will
diminish as $\dot a$ grows, thus rendering the universe flatter.
In inflation, the huge entropy is generated dynamically at the end of
the accelerating phase.

While it is true for a kinetic driven
acceleration that the ratio (\ref{rat}) drops, closer inspection reveals a
subtle kind  of flatness problem for X-branch solutions.
Consider expression (\ref{pm}) with curvature included;
	\begin{equation}
	{d\ln (a^2\p)\over dt}=\pm f{|\dot \p|\over \p}
	\sqrt{1-Z\kappa}
	\label{unnamed}
	\end{equation}
where again
$Z=\left ( a^2\p\over 2C \right )^2$.
For X-branch solutions, the right hand side of
(\ref{unnamed}) is positive and $\Phi a^2$ grows.
Consequently, $Z$ grows and the
curvature term gains in importance relative to
the kinetic term in the square root.
[Although curvature is less important than $H$
(eqn (\ref{rat})),  curvature may
be more important than this piece of $H$.]

Consider the X-branch with different curvatures.
For $\kappa=+1$, the right hand side of (\ref{unnamed})
passes through zero when $Z$ reaches 1.
A branch change is induced as the X-branch
evolves
into
a D-branch solution.
Subsequently, the right hand side of (\ref{unnamed})
is negative, $Z$ drops, and $a^2\Phi$ drops forever.
The usual graceful exit problem of the
D-branch solutions is encountered.

To avoid
this fate, one could require that inflation ends and the
entropy is released before $Z$ gets near 1.
However, this requirement amounts to a kind of flatness problem.
In order to see this, consider
the sufficient inflation  condition (\ref{full}).
We make two assumptions.  Firstly, it is assumed that at some time
$t_{end}$ entropy is produced and inflation ends.
Secondly,  it is
assumed that  the universe has evolved adiabatically since the
end of inflation so that $a_{end}T_{end}=a_oT_o$.
The Hubble constant today can be expressed as
$H_o=\alpha_o^{1/2}T^2_o/M_o$ where, again,  $M_o=1.2 \times 10^{19}$ GeV
is the Planck mass today and
$\alpha_o=\gamma(t_o)\eta_o=(8\pi/3)(\pi^2/30)g_*(t_o)\eta_o$
where
$\eta_o\sim 10^{4}-10^5$
is the
ratio today of the energy density in matter to
that in radiation.
In a model of kinetic inflation  the constraint
(\ref{full}) becomes
	\begin{equation}
	{a_{end}\over a_*}\gta \alpha_o^{-1/2}{M_o\over T_o}
	\left [{d_{\gamma *}^{-1}\over T_{end}}
	\right ] \ \ ,
	\label{cuzit}
	\end{equation}
If $Z<1$ then $a_{end}<{C^{1/2}\over \p^{1/2}_{end}}$.
Notice that for X-branch solutions,
$a_*\Phi_*^{1/2}<a_{end}\Phi_{end}^{1/2}$.
Using these inequalities in the sufficient
inflation condition, along with eqn (\ref{fug}),
gives (with $\delta=0$ for simplicity)
	\begin{equation}
	C^{1/2}\gta 10^{30} {\Phi_{end}^{1/2}\over T_{end}}
	\end{equation}
If the Planck mass at the end of inflation
is $\sim M_o$ and $T_{end}$ is presumably much less
than this, the above condition can only be satisfied
if $C$ is huge;  at least $C\gta 10^{60}$.
This represents a kind of fine tuning, a kind of flatness
problem.

In the Einstein frame,
the flatness problem is seen clearly.
Curvature gains in importance in that picture
since there is no acceleration;
$\tilde R_{curv}^{-1}/\tilde H=|\kappa|^{1/2}/(d\tilde a/d\tilde t)$.
The Einstein expansion, $d\tilde a /d\tilde t$,
is positive and  drops for the X-branch.
As the universe decelerates, curvature becomes
more important.  In order to tilt the energy balance
in favor of the kinetic energy density in the
$\Psi$ field, a large value of $C$ is needed.
This is analogous to requiring a large value of
$S$ in the standard cosmology to tilt the energy balance
in favor
of the radiation energy density.

\section{Discussion}

Conformally related pictures
represent no more than different interpretations
of the same universe.
While these interpretations may disagree wildly,
the results of all measurements in terms of a given
observer's rulers and clocks must be the same.
In the simplest $\Phi{\cal R}$ model, the gravity-driven
acceleration of the cosmic expansion can be
absorbed under a conformal transformation to the
Einstein frame.  An acceleration
of the cosmic expansion in one frame and a
deceleration in a conformally related frame can
be attributed to different interpretations of the
laws of physics and does not represent an inconsistency
in the JBD and the Einstein frame.

However, the condition of sufficient inflation
imposes an additional requirement.  Implicit in
this condition is the demand that today's universe
can be reproduced.  It is easiest
to view inflation upon conformal transformation
to the Einstein frame.  By doing so, it was shown
that the demands of sufficient inflation require
an acceleration of the scale
factor in the Einstein frame.  Since no source
for an acceleration of the Einstein expansion exists,
this argues that a kinetic
inflation alone will be unable to lead to our
smooth observable universe today.

It must be that any obstacle to completing inflation can
be understood entirely in the JBD frame, without reference
to the Einstein frame.
This is in fact the case.
For all D-branch
solutions, the obstacle is manifest in a
graceful exit problem.
This quandry was found in Ref   \cite{ram}
for the specific case of $\om=-1$ which
describes the low energy effective
action of string theory.  It is argued
in this paper that for general $\om(\p)$,
the D-branch always encounters a similar graceful
exit problem.

For X-branch solutions, a general no-go has
been more difficult to formulate directly in the
JBD frame.  For
$\kappa=+1$, the obstacle does surface in general.
For $\kappa=+1$, X-branch solutions evolve into D-branch solutions.
As for all D-branch solutions, graceful exit
is a problem.  A fine tuning of the arbitrary constants
is needed to avoid the induced branch
change.  Such fine tuning was shown to be akin
to the flatness problem.  It is precisely such
unnatural tuning which inflation tries to avoid.
For $\kappa=-1$ and $\kappa=0$,
a no-go in the JBD frame is not immediately obvious.
However, the results
of the Einstein frame predict that any such
attempt will be thwarted.
It can thus be
conjectured that similar obstacles would apply.

The results here indicate the woes of the simplest
attempt.
A kinetic-driven inflation is not
ruled out.  Modifications which lead to a kinetic
acceleration which cannot be transformed away
would be exempt from the arguments in this paper.
An attempt to obviate these concerns
points, for example, to higher order curvature terms or additional
effects from higher order string corrections.
Perhaps with some tenacity, successful kinetic
inflation can be executed.

\centerline{\bf Acknowledgements}

I extend special gratitude to Katie Freese for
bringing the string theory work to my attention.
Thank you also  to Dick Bond, Ram Brustein, Neil Cornish and
Glenn Starkman for their past and/or
current contributions to this work.
I am also grateful for the additional support of the
Jeffrey L. Bishop Fellowship.

\newpage
\centerline{Appendix A: Constraining $\om(\p)$}

A much weaker condition than that of
sufficient inflation can be used to severely
restrict the range of $\om(\p)$ pertinent to inflation.
In Ref \cite{me} it was shown that $\om$ must be negative
and the Planck mass must decrease in order for the
acceleration to be even nominally relevant for inflation.
That discussion is reproduced in brief here.
A nominal condition for the acceleration
to be relevant for inflation is simply that
        \begin{equation}
        d_\gamma>H^{-1}
        \ \ .
        \label{req}
        \end{equation}
If this condition is
not met, then  the scales affected   during
the acceleration were never causally connected.
This is much weaker than the sufficient inflation
condition (\ref{full}).

Use can be made of the flat space results of \S \ref{aqua}, eqn
(\ref{useful}) and (\ref{gam}).
When $\p$ grows, the condition $H>0$ can only be met
when $f>1$ and the upper sign in eqn (\ref{useful}) holds.
For growing $\p$ eqn (\ref{req}) becomes
	\begin{equation}
	f<-{(1-\delta)\over (1+\delta)}
	\label{grob}
	\end{equation}
where $\delta<1$.
Since $f=(1+2\om/3)^{1/2}$ is always positive,
condition (\ref{grob}) is impossible to meet
if $\p$ grows.  Therefore, accelerations
driven by a growing $\p$ cannot be
relevant for inflation.

When $\p$ drops and $f<1$,
both branches are allowed in eqn (\ref{pm}).
Using the above two expressions
eqn (\ref{req})
becomes
        \begin{equation}
        f<\pm \left ( {1-\delta\over 1+\delta}\right )
        \label{sever}
        \end{equation}
The upper sign again corresponds to X-branch
solutions while the lower sign corresponds to
D-branch solutions.
For the X-branch, $\delta <1$
and for the D-branch, $\delta >1$.
The weakest requirement of
(\ref{sever}) is that $f$ be less than 1.
If $f<1$, then $\om <0$.
The
acceleration can therefore only be relevant for
inflation if $\p$ drops and
$\om<0 $.
Although negative $\om$ leads to a kinetic term
in the action with the wrong sign, the net kinetic
energy density can still be positive.
Classically, the kinetic energy density is
positive in an FRW metric if the kinetic coupling
is $\om\ge -3/2$.
The
energy density must be positive in order to banish ghost modes.


\newpage

\centerline{\bf Appendix B:  Sufficient Inflation Revisited}

It is instructive to pursue the sufficient inflation
condition.
The material presented in this appendix
in no way circumvents the trauma discussed in the
body of the paper.  It is only intended to
lend some intuition for the demands
made of the scale factor, the Planck mass
and the heating mechanism.
It is implicit in the requirements of successful
inflation that today's universe results at the end
of the day.
While the left hand side and the
right hand side of eqn (\ref{full}) can be compared
with sheer brute force, that alone does not gaurantee that our universe
results. We will be reminded of this as we come to it.

The sufficient inflation  condition (\ref{full}) can
be expressed as a condition on the growth of the scale factor.
As in \S  \ref{flat} two assumptions are made.
Firstly, it is assumed that at some time
$t_{end}$ entropy is produced and inflation ends.
Secondly,  it is
assumed that  the universe has evolved adiabatically since the
end of inflation so that $a_{end}T_{end}=a_oT_o$.
As before, the Hubble constant today can be expressed as
$H_o=\alpha_o^{1/2}T^2_o/M_o$ where
$\alpha_o=\gamma(t_o)\eta_o=(8\pi/3)(\pi^2/30)g_*(t_o)\eta_o$
and
$\eta_o\sim 10^{4}-10^5$
is the
ratio today of the energy density in matter to
that in radiation.
Finally, using
(\ref{gam}) for $d_{\gamma *}$,
in a model of kinetic inflation  the constraint
(\ref{full}) becomes
	\begin{equation}
	{a_{end}\over a_*}\gta \alpha_o^{-1/2}{M_o\over T_o}
	\left [{| {f \dot \p / \p}(1-\delta)^{-1}|_*\over T_{end}}
	\right ] \ \ .
	\label{cousinit}
	\end{equation}
The flat space results have been used since curvature
cannot aid in a solution to the horizon problem.  In
fact, curvature leads to a problem of its own, namely the flatness
problem (see \S \ref{flat}).
The amount of inflation needed depends on the efficiency
in converting the kinetic energy into temperature
and thus on the specifics of a heating mechanism.

Eqn  (\ref{cousinit}) can be pushed further if some simple
conjectures about the heating mechanism are made.
Let $T_{end}$ be given by
	\begin{equation}
	T_{end}=\epsilon E_{end}
	\end{equation}
where $\epsilon$ is the efficiency with which the kinetic energy density
is converted into entropy and $E_{end}$ is the net
available kinetic energy density.
Suppose the energy available for conversion
into particles is the full energy density in the $\p$ field
times a unit volume,
$E_{end}=\rho_{end}^{\p}a_{end}^3$.
In an FRW cosmology, the energy density in the $\Phi$-field
can be expressed as
        \begin{equation}
        \rho_\Phi={3\over 32\pi}{\dot \Phi^2\over \Phi}
        \left ( f\pm 1\right )^2\ge 0
        \ \
        \label{good}
        \end{equation}
\cite{me}.  So,
$T_{end}$ becomes
	\begin{equation}
	T_{end}=\epsilon \left (
	{C^2\over a_{end}^3\p_{end}}{(f_{end}\pm1)^2\over f_{end}^2}
	\right )
	\ \ .
	\end{equation}
Using this input into the sufficient inflation conditon
(\ref{cousinit}), along with eqn (\ref{fug}) gives the condition
	\begin{equation}
	{a_*^2\ \ \p_*\over a_{end}^2\p_{end}}
	\gta \alpha_o^{-1/2}{M_o\over T_o}{(1-\delta)^{-1} }{f^2\over
	(f\pm 1)^2}{1\over \epsilon C}\ \ .
	\label{ho}
	\end{equation}
Take $(1-\delta)\sim {\cal O}(1)$ and $C\sim{\cal O}(1)$
and $f_{end}\sim {\cal O}(1)$ for  now.
Using $M_o=1.2 \times 10^{19}$ GeV and
$T_o=2.3 \times 10^{13}$ GeV, eqn (\ref{ho})
becomes
roughly
	\begin{equation}
	{a_*^2\ \ \p_*\over a_{end}^2\p_{end}}
	\gta
	{10^{30}\over \epsilon }\ \ .
	\label{hope}
	\end{equation}
Even if the efficiency $\epsilon$ is 1, condition
(\ref{hope}) would require $a^2\p$ to be bigger in the past.
{}From eqn (\ref{pm}), only for D-branch
solutions will $a^2\Phi$ decrease.  For the above
choice of parameters, it follows that only D-branch
solutions will satiate the requirements of sufficient
inflation.\footnote{For D-branch solution, $\delta >1$.
If $\delta_*$ is very large, then, using the definition
of $\delta$, eqn (\ref{ho}) becomes
	\begin{equation}
	{a_i^2\ \ \p_i\over a_{end}^2\p_{end}}
	\gta \alpha_o^{-1/2}{M_o\over T_o}{f_{end}^2\over
	(f_{end}\pm 1)^2}{1\over \epsilon C}\ \ .
	\label{hoboy}
	\end{equation}
where subscript $i$ denotes initial values.
A large $\delta$ means the onset of inflation
is postponed until $a^2\Phi$ drops
substantially from its initial value.
It does not make sufficient inflation
any easier to satisfy.}

If the universe always expands then $a_{end}>a_*$.
This simple inequality can be used in conjunction
with (\ref{hope}) to estimate
the minimum change in the Planck mass,
	\begin{equation}
	{m_{pl}(t_{end})\over
	m_{pl}(t_*)}\lta 10^{-15} \epsilon^{1/2}
	\ \ .
	\end{equation}
The Planck    mass must drop by roughly 15 orders of
magnitude, or more, during inflation.
It is not enough that this condition is satisfied.
As discussed in \S \ref{gracie}, a mechanism
is needed to induce a branch change.  If no branch
change is induced, inspection of the equations
of motion indicate that $H$ would likely pass
through zero and the universe would enter
an age of collapse.
Thus the D-branch would not lead to our expanding
universe today.

Consider instead a different choice of parameters.
For instance $f_{end}=10^{-15}$
so that $\om$ is near $-3/2$ to 1 part in $10^{15}$.
Keep $(1-\delta)\sim {\cal O}(1)$ and $C\sim{\cal O}(1)$.
Then
	\begin{equation}
	{a_*^2\ \ \p_*\over a_{end}^2\p_{end}}
	\gta
	{1 \over \epsilon }\ \
	\label{hoo}
	\end{equation}
and $a^2\Phi$ need not drop.
Thus if $f_{end}$ is fantastically small,
then X-branch solutions may be able to meet condition
(\ref{full}).
Alternatively, $C$ would need to be unnaturally large.
In order for X-branch solutions to address the sufficient
inflation condition, it appears some fine tuning would
be involved.

Before ending, one last comment can be made.
To take a different perspective, the arguments
in this paper reveal that kinetic driven acceleration
alone is not enough to remedy the initial condition
problems of cosmology.  To some extent this is obvious.
An inflationary model combines the acceleration with
some prescription   for heating the universe.
The black box remains the heating mechanism.
Two possible mechanisms are (1) Hawking-Unruh radiation
generated as a consequence of the accelerated expansion
or (2) particle production generated from
oscillations in the Planck field.  The Planck field could
oscillate as a result of
the changing kinetic coupling.  Just as with oscillations
induced by a potential, the oscillating field can decay into particles.
If the heating mechanism involved
unusual physical processes, as it often does, it might
be possible to introduce the effects needed to
amend the problems enumerated in this paper.
While it is highly
unsatisfying to place such unresolved issues
into a black box, the possibility could not go
without mention.

\end{document}